\documentclass[twocolumn,showpacs,nofootinbib,10pt]{revtex4-1}
\usepackage{color,graphicx}
\usepackage[colorlinks,hyperindex]{hyperref}

\definecolor{green1}{RGB}{0,128,0} 
\hypersetup{hidelinks,backref=true,pagebackref=true,hyperindex=true,colorlinks=true,breaklinks=true,urlcolor= blue}
\hypersetup{%
  colorlinks = true,
  linkcolor  = blue,
  citecolor = green1,
}
\usepackage{bookmark}
\usepackage{hyperref,color,xcolor}
\hypersetup{hidelinks,hyperindex=true,colorlinks=true,breaklinks=true,urlcolor= blue}
\hypersetup{%
  colorlinks = true,
  linkcolor  = blue
}

\usepackage{float}\usepackage[all]{xy}
\usepackage{amsmath,upgreek}
\usepackage{amssymb}

\newcommand{\beq}{\begin{eqnarray}}
\newcommand{\eeq}{\end{eqnarray}}
\newcommand{\be}{\begin{equation}}
\newcommand{\ee}{\end{equation}}
\newcommand{\bea}{\begin{eqnarray}}
\newcommand{\eea}{\end{eqnarray}}

\newcommand{\ba}{\begin{eqnarray}}
\newcommand{\ea}{\end{eqnarray}}
\usepackage{tensor}

\begin{document}
\title{The extended minimal geometric deformation of SU($N$) dark glueball condensates}

\author{A. Fernandes--Silva}
\email{armando.silva@ufabc.edu.br}
\affiliation{CCNH, Universidade Federal do ABC - UFABC, 09210-580, Santo Andr\'e, Brazil.}

\author{A. J. Ferreira--Martins}
\email{andre.juan@ufabc.edu.br}
\affiliation{CCNH, Universidade Federal do ABC - UFABC, 09210-580, Santo Andr\'e, Brazil.}

\author{R. da Rocha}
\email{roldao.rocha@ufabc.edu.br}
\affiliation{CMCC, Universidade Federal do ABC -- UFABC, 09210-580, Santo Andr\'e, Brazil.}



\begin{abstract} 

The extended minimal geometric deformation (EMGD) procedure, in the holographic membrane paradigm, is employed to model  stellar distributions that arise upon    
self-interacting scalar glueball dark matter condensation. Such scalar glueballs are SU($N$) Yang-Mills hidden sectors beyond the Standard Model. Then, corrections to the gravitational wave radiation,  emitted by SU($N$) EMGD dark glueball stars mergers, are derived, and their respective spectra are studied in the EMGD framework, 
due to a phenomenological brane tension with finite value. The bulk Weyl fluid that drives the EMGD is then proposed to be experimentally detected by enhanced windows at the eLISA and LIGO.

\end{abstract}



\maketitle

\section{Introduction}

The membrane paradigm of brane-world scenarios can be deployed into the fluid/gravity correspondence as the low energy regime of AdS/CFT conjecture \cite{Eling:2009sj}, which, briefly, states that a theory of gravity in an anti-de Sitter (AdS) space of $d + 1$ dimensions is dual to a strongly coupled conformal field theory (CFT) -- whose low energy underlying hydrodynamics corresponds to the Navier--Stokes equations -- at the $d$-dimensional border of the AdS space \cite{gub,Maldacena:1997re,Bilic:2015uol}. 
 In the membrane paradigm of the fluid/gravity  correspondence  \cite{hub}, black holes in the brane, as well as black branes and black strings, studied from the viewpoint of fluid membranes \cite{CR3,daRocha:2017cxu} in the long wavelength limit \cite{Ovalle:2017wqi,Pinzani-Fokeeva:2014cka}.  Other successful paradigms are also described in the seminal Refs. \cite{maartens,Antoniadis:1998ig}.

In the membrane paradigm and beyond the general relativity (GR),  the so called method of geometrical deformation (MGD)  
places itself as an important procedure to generate new solutions of the effective Einstein's field equations on the brane \cite{Ovalle:2017wqi,ovalle2007,covalle2,darkstars}, describing  black holes and compact stellar distributions as well, in a 5D bulk Weyl fluid bath \cite{Ovalle:2007bn,Casadio:2015jva}. 
The MGD and its extensions take into account the brane Einstein's field equations \cite{GCGR,CoimbraAraujo:2005es}, where the effective stress-energy tensor has additional terms, in particular regarding the Gauss--Codazzi equations from the bulk stress-energy tensor into the brane \cite{maartens}. Important terms are the bulk dark radiation, the bulk dark pressure, the electric part of the Weyl tensor and quadratic terms on the brane stress-energy tensor. This last one is derived for regimes of energy that are further than the (finite) brane tension in the theory. 
 The MGD and its extensions \cite{CR3} have been recently equipped with experimental, phenomenological, and  observational very precise bounds on the running parameters. MGD gravitational lensing effects were explored in Ref. \cite{Cavalcanti:2016mbe} and the classical tests of GR imposed the most strict  bound on the brane tension in Ref. \cite{Casadio:2015jva}. Besides, the configurational entropy was used to provide account for the critical stellar densities, in the MGD, in full compliance with the Chandrasekhar's critical stellar densities that are extremal points of the system associated informational entropy \cite{Casadio:2016aum}. Besides, MGD black holes analogues were explored in Ref. \cite{daRocha:2017lqj} where  sound waves through  de Laval nozzles in a laboratory were shown to be suitable to derive experimental data regarding the 5D bulk Weyl fluid, being acoustic perturbations indeed the analogue MGD quasinormal modes. MGD black branes was also studied in Ref. \cite{CR5}.

On the other hand, gauge forces,  that are not encompassed by the Yang--Mills fields of the Standard Model,   
might play a prominent role in describing features of the observed universe. Indeed, the Standard Model may present a coupling to hidden sectors ruled by a pure Yang-Mills setup,  in the low energy regime.  Non-Abelian dark forces 
can be then implemented by a (pure) Yang-Mills setup,  with simple gauge group,  confining at the energy scale of the theory with dark gluons. This system is then bound into some spectrum of  dark glueballs, that are colour-neutral, that have mass or the order of the energy scale \cite{Juknevich:2009ji,Forestell:2016qhc,ArkaniHamed:2008qn}. 
The scalar glueball dark matter is then a prime candidate emulating   SU($N$) dark gauge hidden sectors  \cite{Boddy:2014qxa,Boddy:2014yra,Juknevich:2009ji,Soni:2016gzf}, whose cross section is very large. 
When bosons and fermions in the Standard Model  are precluded  to interact with the SU($N$) scalar glueball whatsoever, a Bose--Einstein condensation of glueballs can occur, originating thus compact stellar distributions.  In particular, in the holographic AdS/CFT interpretation, Bose--Einstein condensates of scalar glueballs occupy a relevant place \cite{Hartmann:2012wa}.
  Ref. \cite{Soni:2016gzf} discusses relevant elastic scatterings among glueballs, manifesting their feasibility as a self-interacting dark matter candidate.
Dark SU($N$) glueball stars were studied on fluid E\"otv\"os branes in Ref. \cite{daRocha:2017cxu}, with proposed  experimental signatures provided  by gravitational waves their mergers. 
Hereon,  the extended MGD (EMGD) procedure \cite{CR3} is proposed  to explore physical signatures of SU($N$) dark glueball condensates.

\textcolor{black}{This paper is organised as follows: Sect. II is devoted to introduce the EMGD method, reviewing the MGD one, through the deformation of the radial component of the metric, presenting  solutions for the brane Einstein's field equations. The EMGD is implemented by taking into account the deformation of the metric temporal component, given by a $k$ parameter, phenomenologically  driving the extension of the MGD. The Schwarzschild solution is obtained for $k=0$ and, for  $k>0$, terms up to order $\mathcal{O}(r^{-(k+1)})$ are considered.  In Sect. III, the self-interacting system of glueball dark matter is studied, with a self-interacting scalar glueball potential that induces condensation into SU($N$) dark EMGD stellar distributions. Then,  the corrections to the gravitational wave radiation that is emitted by SU($N$) dark EMGD glueball mergers are derived, being their spectra obtained for  two important cases in the EMGD setup, due to a phenomenological brane tension with finite value, opening a wider range probed by the eLISA and the LIGO. In Sect. IV the concluding remarks are outlined with the perspectives.}

\section{EMGD stellar distributions}

The Minimal Geometric Deformation (MGD) procedure can be realized as a mechanism employed to derive high energy corrections to the general relativity (GR), in such a way that the Einstein's field equations non-linearity does not produce inconsistencies in the solutions. In fact, the AdS/CFT correspondence 
can bind warped extra-dimensional models to 4D theories that are strongly-coupled. According to the membrane paradigm, which has been used to realize the deformation method, our 4D universe is a brane with a finite positive tension (or energy density) $\sigma$ \cite{maartens}. 
Systems with energy $E\ll\sigma$ neither feel the self-gravity effects nor the bulk effects, which allows then the recovery of GR in such regime,  corresponding to an infinitely rigid brane, wherein $\sigma\to\infty$. 
The most strict brane tension bound, $\sigma \gtrsim  3.18\times10^6 \;{\rm MeV^4}$, was obtained in the MGD context  \cite{Casadio:2016aum}. 
  
The study the EMGD, the 5D Einstein Field Equations (EFE) in the bulk 
\begin{equation}
{}^{(5)}G_{AB}= {}^{(5)}T_{AB} \label{5deinstein} 
\end{equation}   must be computed,  where ${}^{(5)}G_{AB}$ denotes the 5D Einstein tensor and ${}^{(5)}T_{AB}$ is the 5D stress-energy tensor. Natural units  $8\pi G=c=1=\hbar$ are used hereon, and $M_{\rm Pl}$ shall denote the 4D Planck mass. Greek indexes $\mu,\nu$ represent  the 4D brane indexes running from $0$ to $3$, whilst $A,B=0,\ldots,4$ denote bulk indexes.

The Gauss--Codazzi equations can be used to implement the projection of Eq. (\ref{5deinstein}) onto the brane \cite{GCGR}, resulting in the effective EFE on the brane. The projected EFE presents high energy corrections evinced by the bulk permeated by a 5D Weyl fluid, whose projection onto the brane yields the electric part of the Weyl tensor $\mathcal{E}_{\mu\nu}$. 
The interaction between the brane and the bulk is strictly gravitational, with no exchange of further fields. Therefore, the projection of the 5D Weyl fluid on the brane encrypts the effects due to 5D gravitons for the linearized case, namely, the Kaluza--Klein modes. Effects provided by $\mathcal{E}_{\mu\nu}$ are therefore considered non-local, since it does not depend on any data on the brane, having the form  \cite{maartens}, 
\begin{eqnarray}
\!\!\!\!\!\!\!\!\mathcal{E}_{\mu\nu}(\sigma^{-1}) \!=\!-6\sigma^{-1}\!\left[ \mathcal{U}\!\left(\!u_\mu u_\nu \!+\! \frac{1}{3}h_{\mu\nu}\!\right) \!+\! \mathit{Q}_{(\mu} u_{\nu)}\!+\!\mathcal{P}_{\mu\nu}\right], \label{A4}
\end{eqnarray}
\noindent where $h_{\mu\nu} = g_{\mu\nu} + u_\mu u_\nu$ is the orthogonal projector to the  $4$-velocity $u^\mu$ onto the
brane, being $g_{\mu\nu}$ the brane metric components. Besides, $\mathcal{U}=-\frac16\sigma\mathcal{E}_{\mu\nu} u^\mu u^\nu$ is the effective energy density; $\mathcal{P}_{\mu\nu}=-\frac16\sigma\left(h_{(\mu}^{\;\rho}h_{\nu)}^{\;\sigma}-\frac13 h^{\rho\sigma}h_{\mu\nu}\right)\mathcal{E}_{\mu\nu}$ is the  effective non-local anisotropic stress-tensor; and the effective non-local energy flux on the brane, $\mathit{Q}_\mu = -\frac16\sigma h^{\;\rho}_{\mu}E_{\rho\nu}u^\nu$, is originated from the bulk free gravitational field. Local corrections are encoded in the tensor:
\begin{eqnarray}
S_{\mu\nu} = \frac{T}{3}T_{\mu\nu}-T_{\mu\kappa}T^\kappa_{\ \nu} + \frac{g_{\mu\nu}}{6} \Big[3T_{\kappa\tau}T^{\kappa\tau} - T^2\Big] \ ,
\end{eqnarray}
\noindent where $T_{\mu\nu}$ is the brane matter stress-tensor, in the MGD setup represented by a perfect fluid:
\begin{equation}
T_{\mu\nu}= (\rho +p)u_\mu u_\nu - p g_{\mu\nu}, \label{perfluid}
\end{equation} 
\noindent with $\rho$ being the fluid density and  $p$ represents  the fluid pressure field. Notice that $S_{\mu\nu}$ is a correction arising from the bulk, quadratic in $T_{\mu\nu}$. The 4D effective EFE then read:
\begin{equation}
G_{\mu\nu}
-T_{\mu\nu}-\mathcal{E}_{\mu\nu}(\sigma^{-1})-\frac{\sigma^{-1}}{4}S_{\mu\nu}=0 . \label{projeinstein}
\end{equation} 
Since $\mathcal{E}_{\mu\nu} \sim \sigma^{-1}$, it is straightforward to notice that in the infinitely rigid brane limit,  $\sigma \rightarrow \infty$, GR is recovered and $G_{\mu\nu} = T_{\mu\nu}$. The brane tension has such important role in the MGD, which provides a way to check that the deformation chosen in the method recovers the standard GR equations.

Compact stars in 4D, which must be solutions of Eq. \eqref{projeinstein}, can be described by a static, spherically symmetric metric, written in Schwarzschild-like coordinates as:
\begin{equation}
 ds^2 = -e^{\upnu(r)}\mathrm{d}t^2 + e^{\lambda(r)}\mathrm{d}r^2 + r^2\mathrm{d}\Omega^2 \ ,
 \label{eq:metric_general_spher_symm}
\end{equation}
By solving Eq. \eqref{5deinstein}, using the metric of Eq. \eqref{eq:metric_general_spher_symm} and the stress-energy tensor  \eqref{perfluid}, one obtains the effective pressure components of the Weyl fluid:
\beq
\breve{p}_a &=&  p + \sigma^{-1}\Big(2 \mathcal{U} +(-3+7^a) \mathcal{P} + p\rho+\frac{\rho^2}{2}  \Big) \ , \label{A11}
\eeq
\noindent where either $a=0$, counting the temporal component, or $a=1$, if the radial component is taken into account, and  where $\mathcal{P}=\mathcal{P}_{\mu}^{\;\,\mu}$. Effects of the 5D Weyl fluid makes the fluid projected on the brane to present a certain degree of  anisotropy \cite{ovalle2007,covalle2,darkstars}, encoded into the equation:  
\be
\breve{p}_0 - \breve{p}_1 + {6 \mathcal{P}}{\sigma^{-1}}=0.\label{A6}
\ee
\noindent Eq. (\ref{A6}) asserts that the difference between the components of the fluid pressure is a function of $\sigma^{-1}$, being the fluid isotropic when the GR limit $\sigma^{-1}\to0$ is recovered. 

Considering the inner region of a stellar distribution and the outer vacuum, the radial coordinate $r$ ranges from the center of the star ($r=0$) to the star surface ($r=R$), and then beyond into the outer vacuum, where $\rho = p = 0$ in Eq. (\ref{perfluid}). We shall encode such a condition into the following equation, which is the general standard GR expression for the radial component of the metric:
\begin{equation}\label{999}
 \xi(r) = 
 \begin{cases}
1 - \frac{2M}{r} & \text{ for } r > R, \\ 
 1 - \frac{2m(r)}{r}& \text{ for } r \leq R.
\end{cases}
\end{equation}
\noindent where $m(r)= {8 \pi G_4} \int_0^r \check{r}^2\rho d \check{r}$ stands for the GR mass function. In fact, in the GR limit $1/\sigma \rightarrow 0$, the function  $M=M(\sigma)$ in Eq. (\ref{999}) is led to the standard GR mass value, $M_0$. This  limit yields  $\left .M\right\rvert_{\sigma \rightarrow \infty} = M_0 = m(R)$.

For the metric of Eq. (\ref{eq:metric_general_spher_symm}) to be a solution of Eq. (\ref{projeinstein}), one assumes the \emph{ansatz} for its radial component \cite{covalle2}:
\begin{equation}
 e^{-\lambda(r)} = \xi(r) + f(r) \ ,
 \label{eq:radial_mgd}
\end{equation}
where the geometric deformation given by $f(r)$ is what an observer on the brane experiences due to the projected 5D bulk gravity effects onto the brane. In fact, $f(r)$ is a distorting function added to the standard GR solution, given by $\xi(r)$, therefore yielding a deformation, reading \cite{covalle2}:
\begin{equation}
\begin{gathered}
\!\!\!\!\!\! f(r) \!=\!  e^{-I}\int \frac{ 2 r e^I}{r\partial_r \upnu \!+\! 4} \left ( H \!+\! {(\rho^2 \!+\! 3 \rho p)}{\sigma^{-1}} \right ) dr 
 \!+\!\chi e^{-I},
 \label{eq:geo_def}
 \end{gathered}
\end{equation}
\noindent where one denotes $\partial_r = d/dr$, $H = H(p,\rho, \upnu)$,  $\chi = \chi(\sigma)$ is an integration constant, and
\begin{equation}
 I(r) = \int_{r_0}^r \frac{ 2\check{r}\left (\frac{2}{\check{r}^2}+ \partial_{\check{r}}^2 \upnu + \frac{ \left (\partial_{\check{r}}\upnu \right)^2}{2} + \frac{2 \partial_{\check{r}}\upnu}{\check{r}} \right )}{\check{r}\partial_{\tilde r} \upnu + 4} d\check{r} \ ,
 \label{eq:def_i}
\end{equation}
\noindent with $r_0$ chosen according to the region of interest.

The most important contribution in the geometric deformation comes from the function $H$,
\begin{equation}
\begin{gathered}
H(\upnu, p, \rho) = - \xi \left ( \partial_{r^2}^2 \upnu + \frac{(\partial_r \upnu)^2}{2} + \frac{2 \partial_r \upnu}{r} + \frac{1}{r^2} \right ) +
\\
\qquad\quad+ \frac{1}{r^2} - \partial_r \xi \left ( \frac{\partial_r \upnu}{2} + \frac{1}{r} \right ) + 3 p.
\end{gathered}
\end{equation}
In fact, the field $H$ equals zero for the standard GR solution, yielding an infimum for the MGD, since $H> 0$ \cite{Casadio:2015jva}. Thus, in this limit, the geometrical deformation shall be solely driven by the pressure and the energy density of the source. When it sets in, the MGD, $\check{f}(r)$, is explicitly given by:
\begin{equation}
 \check{f}(r) = \frac{e^{-I}}{\sigma}  \int \left (\frac{2re^I}{r \partial_r \upnu + 4} \right ) (\rho^2 + 3\rho p) dr + \chi e^{-I}.
\end{equation}
Therefore, starting from the choice $\upnu = \upnu_{\rm GR}$, one can derive the deformed radial component of the metric by evaluating Eq. (\ref{eq:radial_mgd}) with the minimal deformation $\check{f}(r)$ \cite{Ovalle:2017wqi}. Afterwards, the deformed temporal metric component is derived with the remaining expressions from the effective EFE relating $\lambda$ and $\upnu$. 

The field $\chi$ in Eq. (\ref{eq:geo_def}) must be zero in the standard GR limit.  Besides, the integration constant $\chi$ must equal to zero in the star inner region ($r<R$), for the metric to be smooth at $r=0$. Notwithstanding, at $r>R$, $\chi$ can assume values that are different of zero. Hence, a geometrical deformation can be associated with the standard GR Schwarzschild (``$S$'') solution: $e^{\upnu_{\rm S}} = e^{-\lambda_{S}} = 1 - \frac{2M}{r}$, in which case the MGD field  becomes $\check{f}(r)\rvert_{ p =\rho = 0} = \chi e^{-I(r)}$, implying \cite{ovalle2007}:
\begin{equation}
\left.\check{f}(r)\right\rvert_{\rho = p = 0} =   \frac{ b \left (1-\frac{2M}{r} \right )}{r \left ( 1-\frac{3M}{2r} \right )}\chi \ ,
\end{equation}
\noindent where $
 b = b(M) \equiv \frac{R\left(1-\frac{3M}{2R}\right)}{1-\frac{2M}{R}}.$ 
 In Ref. \cite{Casadio:2015jva} such a parameter was phenomenologically weaker bound as $|b(M)|\lessapprox5\times 10^{-11}$. 
 
Therefore, the deformed exterior temporal and radial metric components are respectively given by:
\begin{eqnarray}
 e^\upnu &=& 1 - \frac{2M}{r},
 \label{eq:mgd_temp}\\
 e^{-\lambda} &=& \left (1 - \frac{2M}{r} \right ) \left ( 1 + \frac{b\chi}{r \left (1- \frac{3M}{2r} \right )}\right ).
 \label{eq:mgd_rad}
\end{eqnarray}
MGD black holes, described by this metric, has two event horizons, $
r_1 = 2M$ and $r_2 = \frac{3M}{2} - \psi$. Thus, up to the second order on $\sigma^{-2}$, the deformation term in Eq. (\ref{eq:radial_mgd}) yields $
f(r) =  - \frac{\psi}{r}.$ 
For a stellar distribution, 5D bulk effects are maximal, near the MGD stellar distribution surface $r=R$. The parameter $|\psi|$ stands for  the star compactness, being higher for higher values of $\psi$. Observational data imply the strongest bound $|\psi|\lesssim 3.18 \times 10^{-11}$ \cite{Casadio:2015jva}, hence matching good results up to the order ${\cal O}(\sigma^{-2})$.
The MGD procedure was used in Ref. \cite{daRocha:2017cxu} to model dark SU($N$) stars and to study their experimental signatures. Furthermore, another application was accomplished in the context of  de Laval nozzles, which can be associated to MGD black hole analogues, so that 5D effects might be observed in the laboratory \cite{daRocha:2017lqj}. 

Now, an extension of the MGD can be regarded \cite{CR3}, generalizing the MGD for the outer region. It can be accomplished by considering a deformation not only on the radial but also on the temporal metric component. This determines what is called the extended minimal geometric deformation (EMGD) method, which will be discussed in this section, as introduced in \cite{CR3}.

Let us take a geometric deformation of the temporal metric component $\upnu(r) = \upnu_S + h(r)$ in eq. (\ref{eq:metric_general_spher_symm}) \cite{ovalle2007}. The  $\upnu_S$ term defines, as previously, the Schwarzschild temporal metric component, and $h(r)$ is the temporal deformation, which is proportional to $\sigma^{-1}$, thus providing the GR limit. Using the effective vacuum  EFE, the radial geometric deformation $f(r)$ can be written in terms of $h(r)$ as \cite{CR3}:
\begin{equation}
 f(r) = e^{-I} \left ( \chi - \int_R^r \frac{2\check{r}e^{I} G(h)}{\check{r}\partial_{\check{r}}\upnu + 4} d \check{r} \right ) \ ,
 \label{eq:geo_ext}
\end{equation}
\noindent where $I$ is given by Eq. (\ref{eq:def_i}), and now:
\begin{equation}
\begin{gathered}
G(h) \!=\!  \xi \left ( \partial_{r^2}^2h \!+\! \partial_r \upnu_{S}\partial_r h \!+\!  \frac{2\partial_r h}{r}\right ) \\\quad\qquad+\frac{1}{2} \left [ \xi(\partial_r h)^2 \!+\! \partial_r\xi\partial_r h \right ].
\label{eq:G(h)}
\end{gathered}
\end{equation}
Therefore, the exterior deformed radial metric component becomes
\begin{equation}
 e^{-\lambda(r)} =  1 - \frac{2M}{r} + f(r) \ , 
 \label{eq:radial_emgd}
\end{equation}
\noindent with the extended geometric deformation $f(r)$ redefined according to Eq. (\ref{eq:geo_ext}).

Notice that a constant $h$ implies $G=0$, which produces an exterior MGD as one had before. On the other hand, it is also possible to achieve a MGD with a $h(r)$, which is given by setting Eq. (\ref{eq:G(h)}) equal to zero. The solution of the resulting differential equation reads \cite{CR3}:
\begin{equation}
 e^{h / 2} = a + \frac{d}{2M} \left ( 1 - \frac{2M}{r}\right )^{-1/2} .
\end{equation}
With the assumption of asymptotic flatness, $\lim_{r \rightarrow \infty}  e^{\upnu} =1$, implying $h \rightarrow 0$, the integration constants $a$ and $d$, both dependent of the brane tension, are constrained by $a = 1 - \frac{d}{2M}$. Hence, the minimally-deformed temporal metric component becomes \cite{CR3}:
\begin{equation}
\! e^{\upnu(r)} = \left ( 1 \!-\! \frac{2M}{r}\right ) \!\left [1 \!+\! \frac{b(\sigma)}{2M} \left(\!\left (\! 1 \!-\! \frac{2M}{r}\right )^{\!-\!1/2} \!\!\!\!\!\!\!-\! 1 \right )^2\right].
\end{equation}
The minimally-deformed radial metric component, on the other hand, reads:
\begin{equation}
 e^{-\lambda(r)} =  1 - \frac{2M}{r} + \chi e^{-I} .
\end{equation}
A more general solution for the exterior radial metric component of Eq. (\ref{eq:radial_emgd}) can be derived \cite{CR3}, under a geometric deformation such that $G(h) \neq 0$. The choice $h(r) = k \log \left ( 1 - \frac{2M}{r}\right )$ yields:
\begin{equation}
e^\upnu = \left ( 1 - \frac{2M}{r}\right )^{k+1} \ ,
  \label{eq:temp_emgd}
\end{equation}
\noindent where $k$ is a constant known as the deformation parameter. Naturally, $k=0$ results no temporal geometric deformation, being directly associated with the Schwarzschild metric. For $k=1$, one has
\begin{equation}
e^{\upnu(r)} = 1 - \frac{4M}{r} + \frac{4M^2}{r^2} \ ,
\end{equation}
\noindent which then allows the calculation of the radial metric component, through Eq. (\ref{eq:radial_emgd}), which yields \cite{CR3}:
\begin{equation}
 e^{-\lambda(r)} = 1 - \frac{2M - \kappa_1}{r} + \frac{2M^2 - \kappa_1 M}{r^2} \ ,
\end{equation}
\noindent for $\kappa_1 = \dfrac{M\chi}{1 - M/R}$.
Now, in order to the radial metric component asymptotically approach the Schwarzschild behavior with ADM mass $\mathcal{M} = 2M$, $
 e^{-\lambda(r)} \sim 1 - \frac{2\mathcal{M}}{r} + \mathcal{O}(r^{-2})$, one  must necessarily have $\kappa_1 = -2M$. In this case, the temporal and spatial components of the metric shall be inversely equal to each other (as it is the case of the Schwarzschild solution), containing a tidal charge $\mathcal{Q} = 4M^2$ reproducing a  solution that is  tidally charged by the Weyl fluid \cite{dadhich}:
\begin{equation}
 e^\upnu = e^{-\lambda} = 1 - \frac{2\mathcal{M}}{r} + \frac{\mathcal{Q}}{r^2}
 \label{eq:k=1}
\end{equation}
It is worth to emphasize that the metric of Eq. (\ref{eq:k=1}) has a degenerate event horizon at $r_h = 2M = \mathcal{M}$. Since the degenerate horizon lies behind the Schwarzschild event horizon, $r_h = \mathcal{M} < r_s = 2\mathcal{M}$, 5D bulk effects are then responsible for decreasing the gravitational field strength on the brane.

Now the exterior solution for $k=2$ can be constructed, making Eq. (\ref{eq:temp_emgd}) to yield:
\begin{equation}
 e^{\upnu(r)}  = 1 - \frac{2 \mathcal{M}}{r} +\alpha(\mathcal{Q},\mathcal{M})\ ,
\end{equation}
\noindent for 
\begin{equation}\alpha(\mathcal{Q},\mathcal{M})= \frac{\mathcal{Q}}{r^2} - \frac{2\mathcal{Q} \mathcal{M} }{9r^3},
\end{equation} where $\mathcal{Q} = 12M^2$ and $\mathcal{M} = 3M$. 
The radial component, on the other hand, reads:
\begin{widetext}
\begin{equation}
\begin{gathered}
  e^{-\lambda(r)}  = \left ( 1 - \frac{2 \mathcal{M}}{3r}\right )^{-1} \left [ \frac{128 \kappa_2}{r} \left ( 1 - \frac{\mathcal{M}}{6r} \right )^7 + \frac{5}{4644864} \frac{ \mathcal{Q}^4}{r^8} +\frac{5}{82944}\left(8 -\frac{\mathcal{M}}{r}\right)\frac{ \mathcal{Q}^3}{r^6} \right ] +
  \\
   \qquad\quad\qquad\quad\qquad\quad+ \left ( 1 - \frac{2 \mathcal{M}}{3r}\right )^{-1} \left [ \frac{25}{1728}\left(6 - \frac{\mathcal{M}}{r}\right)  \frac{ \mathcal{Q}^2}{r^4} + \frac{5}{12}\left (2 - \frac{\mathcal{M}}{r}\right )\frac{\mathcal{Q}}{r^2} -\frac{4\mathcal{M}}{3r} + 1     \right ] \ ,
\end{gathered}\label{eq:k=2}
\end{equation}
\end{widetext}
\noindent where $\kappa_2 = {R \chi}{(2 - M/R)^{-6}}$. The asymptotic Schwarzschild behavior is then assured when $\kappa_2 = -M/32$. In this case, the degenerate event horizon is at $r_e \approx 1.12 \mathcal{M}$ \cite{CR3}, which makes clear that here as well the 5D bulk effects induce a weaker gravity.

The classical tests of GR applied to the EMGD metric provide the following constraints on the value of the deformation parameter: $k \lesssim 4.5$ for the perihelion precession of Mercury, $k \lesssim 4.3$, for the deflection of light by the Sun,  and $k \lesssim 4.2$ for the gravitational redshift of light. Hence, observational and experimental data from the classical tests of GR make us hereon to opt to study the cases $k=1$ and $k=2$, as above detailed.
\vspace*{-0.3cm}

\section{SU($N$) EMGD glueball dark stellar distributions and their gravitational wave radiation spectra}

Hidden SU($N$) gauge sectors may be described by the (scalar) glueball dark matter parading \cite{Juknevich:2009ji,Soni:2016gzf,Bernardini:2016qit}. In fact,  glueballs can interact by exchanging gravitons, constituting a self-interacting system. The number, $N$, of colors driving the SU($N$) gauge sectors and the scalar  glueball mass, $m$, are the parameters that model the glueballs self-interaction. These two parameters drive the $2\to 2$ elastic scattering cross
sections of the lightest
glueball state in the hidden dark sector, given by $\sigma\sim m^{-2} N^{-4}$ \cite{Soni:2016gzf}. Dark SU($N$) glueball stars can be  formed when the self-gravity 
unbalances the energy density of the system, inducing a Bose-Einstein condensation of the glueball system into stellar distributions. This phenomenon can occur in the parameters ranges $10^3 \lessapprox N \lessapprox 10^6$ and $10$ eV $\lessapprox  m \lessapprox$ 10 KeV \cite{Soni:2016gzf,Forestell:2016qhc}.

The action for the glueball system is given by 
\begin{equation}\label{pot4}
\mathcal{S} =  \int d^4x\,\left(\frac{1}{2} g^{\mu\nu} \partial_\mu \phi \partial_\nu\phi - V(\phi)\right).
\end{equation}
The resulting Klein--Gordon--Einstein system can be then scrutinized in what follows, for a $\phi^4$ self-interacting glueball potential, condensating into an 
EMGD stellar distribution due to a finite brane tension, 
\begin{equation}\label{quartic}
V(\phi) = \frac{m^2}{2} \phi^2 + \frac{\lambda}{24}  \phi^4,
\end{equation}
for $\lambda>0$, making the self-interacting force to  counterbalance the gravitational force. Hence, a stable configuration for a SU($N$) dark stellar distribution can be naturally achieved \cite{Soni:2016yes,Soni:2016gzf,daRocha:2017cxu} and therefore the range of  frequencies for gravitational waves generated from SU($N$) EMGD dark stars mergers, will be derived and studied.   
 The Klein--Gordon--Einstein system, derived from Eqs. (\ref{eq:metric_general_spher_symm}, \ref{pot4}, \ref{quartic}),  with the assumption that spherically symmetric  glueball fields are also  periodic in time, $\phi(r,t) = \Phi(r)\exp(i\omega t)$  \cite{Soni:2016gzf}, read \begin{widetext}
\begin{subequations}
\begin{eqnarray}
\Phi''(x) + \left( \frac12e^{\upnu'(x)-\lambda'(x)} +\frac{2}{x}\right) \Phi'(x) -  \left[ \frac{\uplambda}{2}  \Phi^2(x)+ \left(1-{\Omega^2} e^{\upnu(x)}\right) \right]e^{-\lambda(x)}\Phi(x)  &=&0 \label{kg2} \,\\
48{\upnu'(x)}{e^{\lambda(x)}}   - {6x}{\Phi^{\prime2}(x)}e^{\lambda(x)} +{\uplambda}  x\Phi^4(x)+ \frac{1}{x}\left(e^{\lambda(x)}-1 \right)+ {12 x}\Phi^2(x)  \left(1- {\Omega^2}{e^{\upnu(x)}} \right)&=&0 \ , \\
8\frak{M}'(x) - \left[\frac{\uplambda}{6}  \Phi^4(x) + {\Phi^{\prime2}(x)}{e^{\lambda(x)}}+ \left(1+{\Omega^2}{e^{\upnu(x)}} \right) \Phi^2(x)  \right]{x^2}&=&0 \,,
\end{eqnarray}
\end{subequations}
\end{widetext}
where \cite{Soni:2016yes,Soni:2016gzf} $\frac{\frak{M}(x)}{m}$ represents the star mass within a radius $\frac{x}{m}$, yielding dimensionless $\frak{M}$ and $x$ parameters. Besides, one also uses the notation $\uplambda = \frac{12\lambda}{m^2}$ and $\Omega = \frac{\omega}{m}$.
The glueball (dark) matter is then modelled by Eq. (\ref{kg2}), yielding 
\begin{eqnarray}
&&\left[\upphi''({\rm x}) + \left( \frac{2}{{\rm x}} + \frac12e^{\upnu'(x)-\lambda'(x)} \right) {\upphi}'({\rm x})\right] \nonumber\\&&-\uplambda e^{\lambda(x)}\upphi({\rm x}) \left[ \left(1-{\Omega^2}e^{\upnu({\rm x})}  \right) - \frac{ \upphi^2({\rm x})}{2} \right] =0 \ , \label{Klein1} 
\end{eqnarray} where  ${\rm x} = \frac{x}{\sqrt{\uplambda}}$, $\upphi = \sqrt{2\uplambda} \Phi,$ and 
${\rm M} = \frac{\frak{M}}{\sqrt{\uplambda}}$. 
The $\uplambda\gg1$ limit can induce the  first term in Eq. (\ref{Klein1}) to be dismissed, yielding   $
\lim_{\uplambda\gg1}\upphi({\rm x}) = \sqrt{2({\Omega^2}e^{\upnu({\rm x})} - 1})$, implying  that 
\begin{eqnarray}
\!\!\!\!&&\!\!\!\!{\rm M}'({\rm x}) \!-\! {\rm x}^2 \left[ \frac{1}{4} \left( {\Omega^2}e^{\lambda({\rm x})} \!+\! 1 \right) \upphi^2({\rm x}) \!+\! \frac{3}{32} \upphi^4({\rm x}) \right]=0 \ , \label{einstein1} \\
\!\!\!\!&&\!\!\!\!2e^{\lambda({\rm x})}\upnu'{\rm x}^2 -4 {\rm M}+
\left[\left(1\!-\! {\Omega^2}e^{\upnu}\right){\upphi^2} \!+\! \frac{3\upphi^4}{8}\right]{\rm x}^3=0. \label{einstein2}
\end{eqnarray}  

 Figs.~1 -- 3 depict the numerical computations,  where below 
 we consecutively present the results for the EMGD, for $k=1$ and $k=2$, comparing with the standard MGD ($k=0$) in Eq. (\ref{eq:temp_emgd}), and its subsequent GR limit.  
 In what follows,  the brane tension value $\sigma \sim 10^6 \;{\rm MeV^4}$ shall denote the exact current bound $\sigma \gtrsim  3.18\times10^{-6} \;{\rm GeV^4}$ derived in Ref. \cite{Casadio:2016aum}. 
  \begin{figure}[H]\label{1144}
\centering\includegraphics[width=8.9cm]{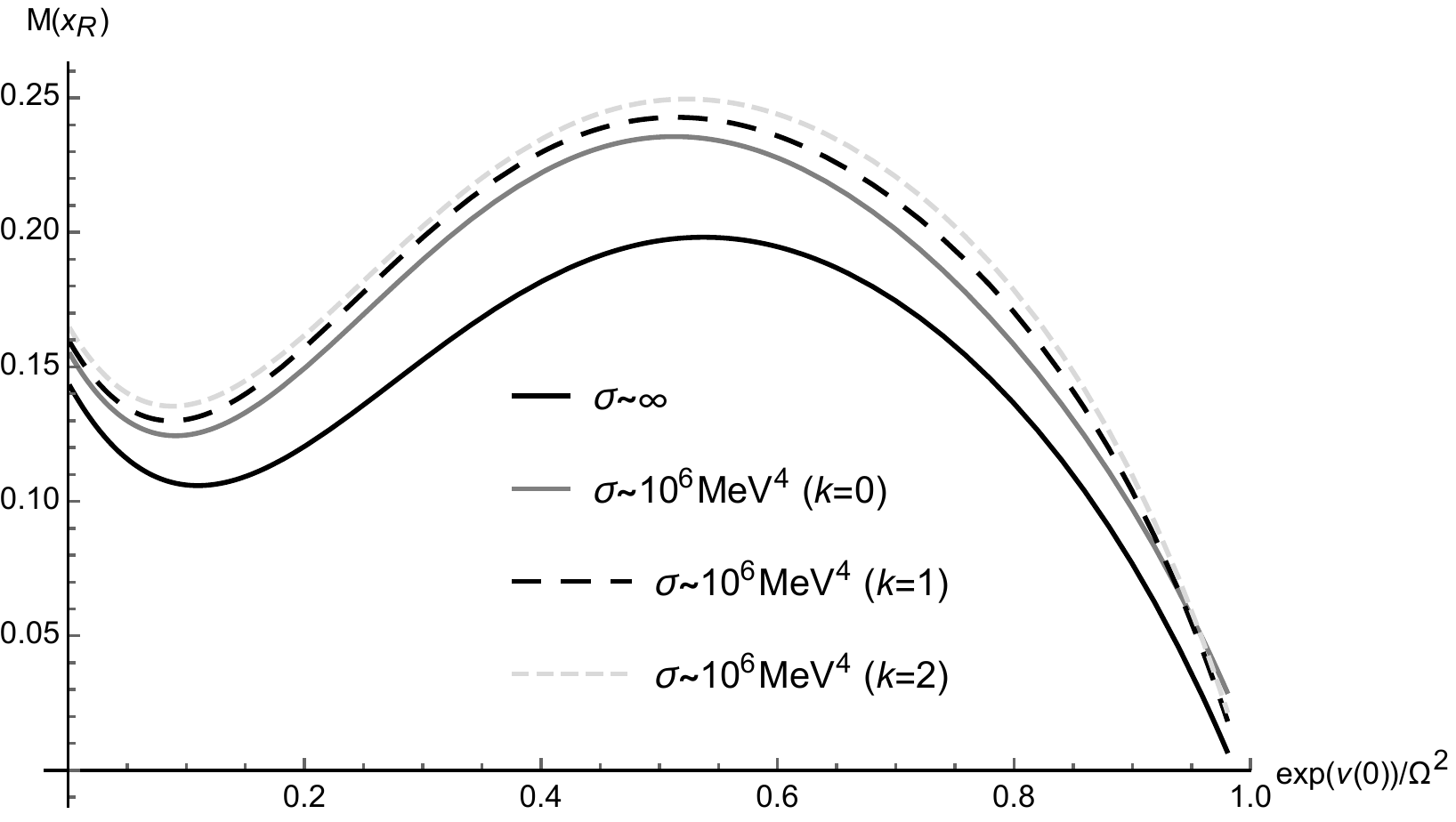}
\caption{\footnotesize Dark SU($N$) EMGD star mass ${\rm M}(x_R)$,  normalized by $\frac{\sqrt{2\lambda}\,M_{\rm pl}^3}{m^2}$, with respect to $\frac{e^{\upnu(0)}}{\Omega^2}$. The black line regards the GR limit $\sigma\to\infty$; the other lines depict the phenomenological bound for the brane tension $\sigma \approxeq 3.18\times 10^{-6}$ GeV$^4$: the standard MGD procedure is a particular case of the EMGD for $k=0$ (gray line); the EMGD case is plot for $k=1$ (black dashed line) and for $k=2$ (gray dashed line).}
\end{figure}\begin{figure}[H]
\centering\includegraphics[width=7.9cm]{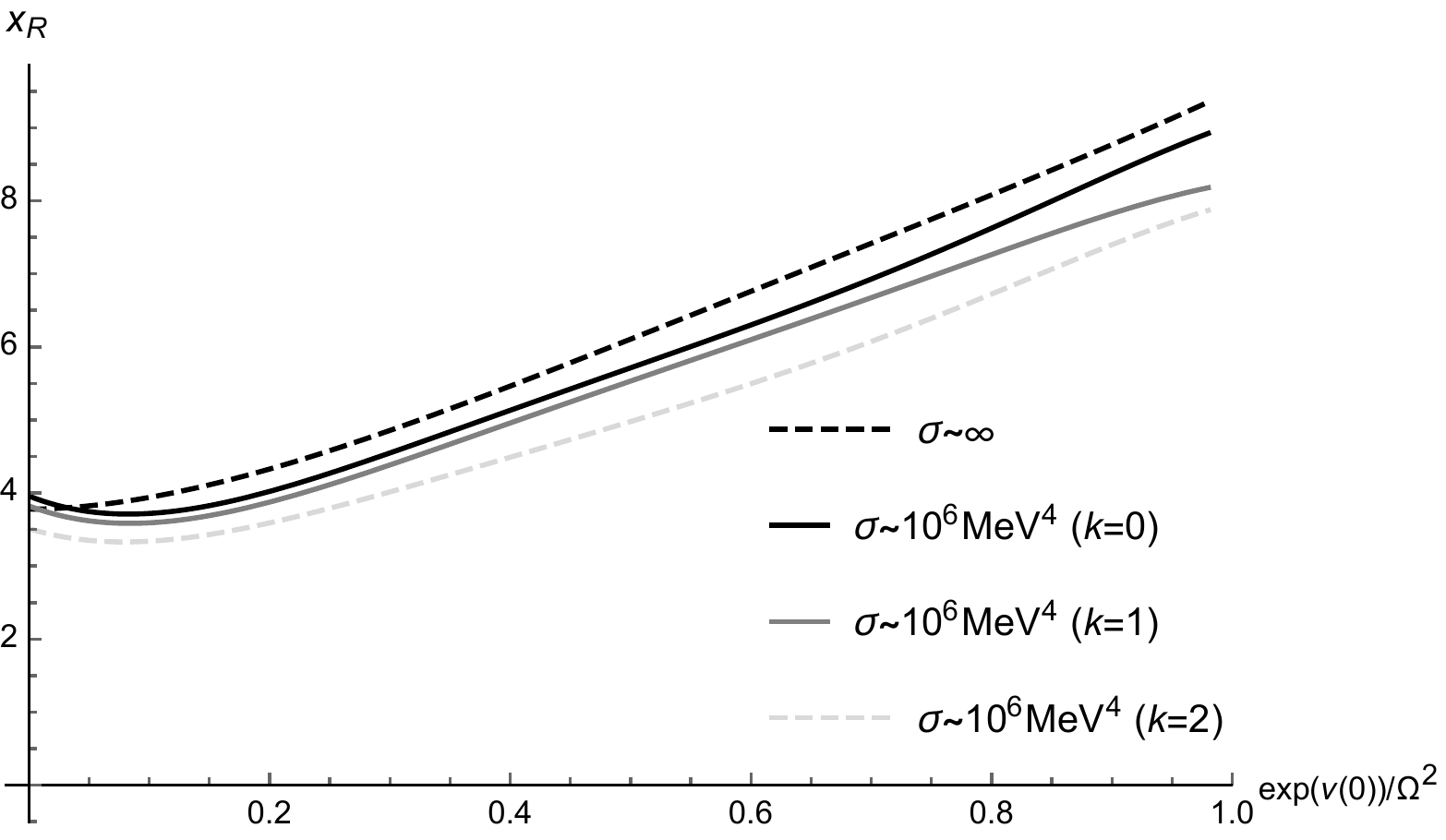}
\caption{\footnotesize Dark SU($N$) extended MGD star radius $x_R$, normalized by $\frac{\sqrt{2\lambda}\,M_{\rm pl}}{m^2}$, with respect to $\frac{e^{\upnu(0)}}{\Omega^2}$. The dashed black line regards the GR limit $\sigma\to\infty$; the other lines depict the phenomenological bound for the brane tension $\sigma \approxeq 3.18\times 10^{-6}$ GeV$^4$: the standard MGD procedure is a particular case of the EMGD for $k=0$ (black line); the EMGD case is plot for $k=1$ (gray line) and for $k=2$ (gray dashed line).}
\end{figure}
 \begin{figure}[H]
\centering
\includegraphics[width=8.9cm]{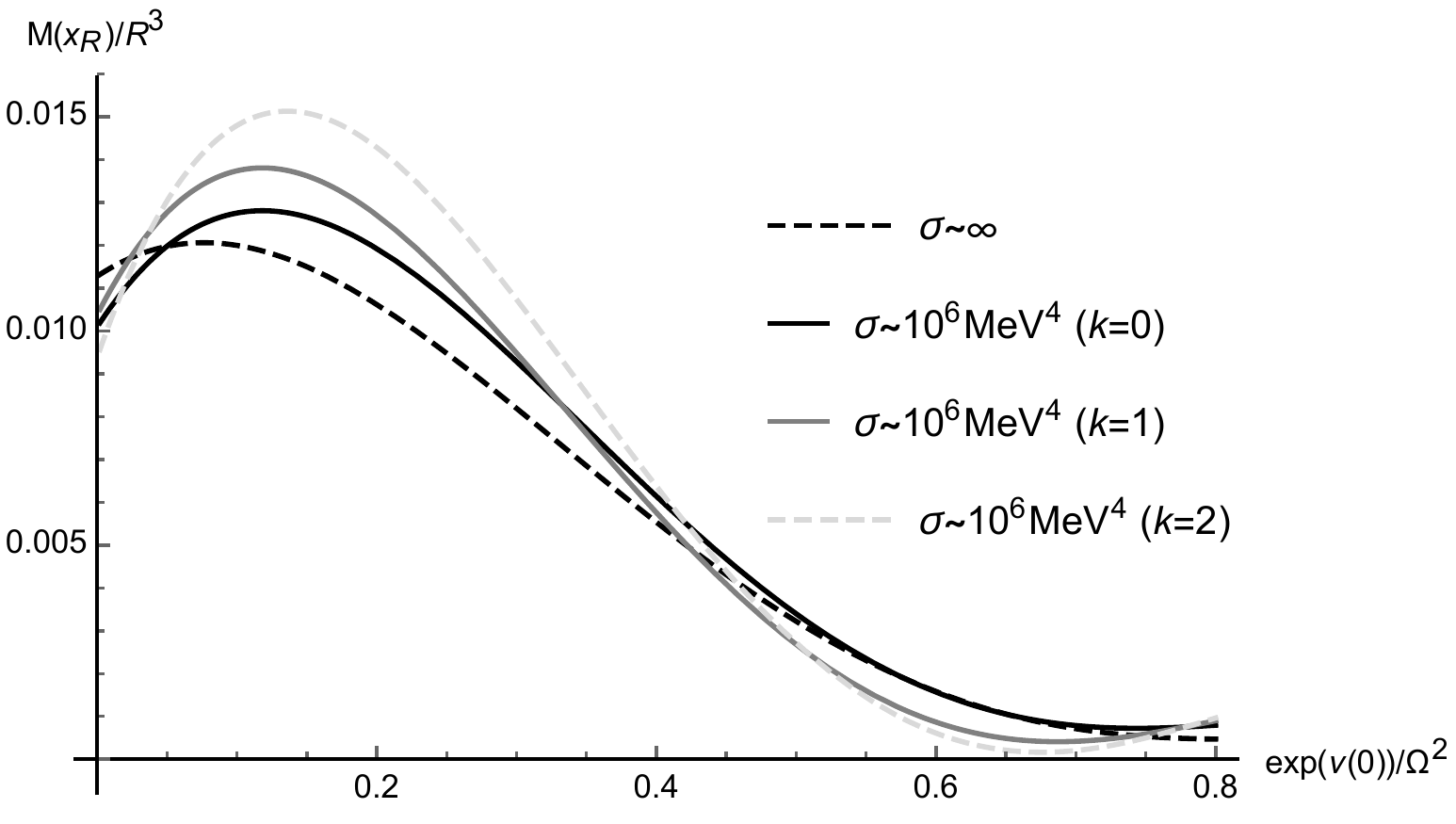}
\caption{Dark SU($N$) MGD star ratio $\frac{{\rm M}(x_R)}{x_R^3}$, normalized by $\frac{\sqrt{2\lambda}\,M_{\rm pl}}{m^2}$, with respect to  $\frac{e^{\upnu(0)}}{\Omega^2}$. The dashed black line regards the GR limit $\sigma\to\infty$; the other lines depict the phenomenological bound for the brane tension $\sigma \approxeq 3.18\times 10^{-6}$ GeV$^4$. The standard MGD procedure is a particular case of the EMGD for $k=0$ (black line); the EMGD case is plot for $k=1$ (gray line) and for $k=2$ (gray dashed line).}\label{133}
\end{figure}

Fig. 1 shows that a SU($N$) EMGD dark stellar distribution can accrete the  dark matter that surrounds it \cite{Soni:2016yes}, having the stellar mass  increased up to a maximal mass, shown  in Table I. 
\begin{center}
\begin{table}[!h]
\begin{tabular}{||c||c|c|c|c|||}\hline\hline
\;Brane tension $\sigma$\vspace{0.06cm}\;&\;$
\frac{e^{\upnu(0)}}{\Omega^2}$\;& $\frac{{\rm M}(x_R)}{x_R^3}$\\\hline\hline
$\infty$ (GR limit)&0.533&0.090\\\hline
$10^{6}$  MeV$^4$ ($k=0$)&0.509&0.142\\\hline
$10^{6}$  MeV$^4$ ($k=1$) &0.518&0.153\\
\hline
$10^{6}$  MeV$^4$ ($k=2$)&0.537&0.172\\\hline
\hline
\end{tabular}\caption{\footnotesize Peak values of the dark SU($N$) EMGD stars radius $\left(\text{normalized by $\frac{\sqrt{2\lambda}}{m^2}\,M_{\rm pl}$}\right)$ and mass $\left(\text{normalized by $\frac{\sqrt{2\lambda}}{m^2}\,M_{\rm pl}^3$}\right)$, as functions of the fluid brane tension, for the MGD ($k=0$ and for the EMGD ($k=1$ and $k=2$).}
\end{table}
\end{center}
As  general SU($N$) dark glueball stellar distributions have typical mass and radius respectively given by  \cite{Soni:2016yes}, 
\begin{eqnarray}\label{massa}
M &=&  \frac{\sqrt{2\lambda}}{m^2}\,M_{\rm pl}^3\, {\rm M}(x_R),\\
R &=&   \frac{\sqrt{2\lambda}}{m^2} M_{\rm pl}\,x_R,\label{raioo}
\end{eqnarray}
the glueball dark SU($N$) EMGD dark star have, then, the following values for its mass and radius,
\begin{eqnarray}\label{massa1}
\!\!\!\!\!\!\!\!\!\!\!\!\!\!\!\!\!\!\!\!\!\!\!\!\!\!\!\!\!\!\!\!\!\!\!\!\!R\!= \begin{cases}902.5 \;m^2 \sqrt{\lambda},&\!\!\quad \text{for $\sigma\to\infty$ (GR limit)},\label{mm1}\\
922.5 \;m^2 \sqrt{\lambda},&\!\!\quad \text{for $k=0$},\label{mfka4}\\
951.1 \;m^2 \sqrt{\lambda},&\!\!\quad \text{for $k=1$},\label{mfka4}\\
982.8 \;m^2 \sqrt{\lambda},&\!\!\quad \text{for $k=2$},\label{mfka4}
\end{cases}\\
\!\!\!\!\!\!\!\!\!\!\!\!M\!= \begin{cases}\frac{9\sqrt{\lambda}}{m^2}10^{-2}M_{\odot},&\!\!\quad \text{for $\sigma\to\infty$ (GR limit),}\label{mm1}\\
\frac{10.94\sqrt{\lambda}}{m^2}10^{-2}M_{\odot}&\text{for $k=0$},\label{mfka4}\\
\frac{12.91\sqrt{\lambda}}{m^2}10^{-2}M_{\odot}&\text{for $k=1$},\label{mfka4}\\
\frac{14.23\sqrt{\lambda}}{m^2}10^{-2}M_{\odot}&\text{for $k=2$},\label{mfka4}
\end{cases}\label{raioo1}\end{eqnarray} 
where $M_{\odot}$ denotes, as usual,  the Solar mass.

The range of frequencies emitted from SU($N$) (Schwarzschild) dark mergers read $f_{\rm max} =\frac{1}{2\pi}\left(\frac{GM}{R^3}\right)^{1/2}$ \cite{Soni:2016gzf,Abbott:2007kv}. SU($N$) EMGD dark stars mergers represent candidates to enlarge such range of frequencies, as we shall show in what follows. Eqs.~(\ref{massa1}, \ref{raioo1}), and Table I, thus imply that the peak  gravitational wave frequency reads
\begin{eqnarray}\label{aaaa}
\!\!\!\!\!\!\!\!\!\!\!\!\!\!\!f_{\rm max} = \sqrt{\frac{\lambda}{2}}\frac{m^2}{\pi M_{\rm pl}}  \!\left(\frac{{\rm M}(x_R)}{x_R^3}\right)\!\Big{\vert_{\rm max}}\approxeq
50 \gamma(k,\sigma)  {\rm Hz},
\end{eqnarray} 
where the function $\gamma(k,\sigma)=  4m^2\sqrt\lambda\;\mathfrak{c}\;10^4/ {\rm GeV}^{2}$ is represented by a factor 
\begin{eqnarray}\label{mfka1}\mathfrak{c} =\mathfrak{c}(k,\sigma)=
\begin{cases}1,&\quad \text{for $\sigma\to\infty$ (GR limit)},\label{mfka1}\\
1.262,&\quad \text{for $k=0$,}\label{mfka4}\\
1.521,&\quad \text{for $k=1$,}\label{mfka4}\\
1.960,&\quad \text{for $k=2$}\label{mfka4}.
\end{cases}\end{eqnarray}
that measures the corrections further the unit, which regards the $\sigma\to\infty$ GR limit, for different values of the brane tension. 
Such a parameter $\mathfrak{c}$ hence provides corrections to peak wave frequencies from SU($N$) EMGD dark star mergers, in a finite tension membrane paradigm.

The range of gravitational wave frequencies is clearly a function that depends on the $m$ glueball dark matter mass and on the number $N$ of colors driving the hidden gauge sectors, and may be detected by the eLISA and the LIGO experiments \cite{Abbott:2007kv}. The relevant stars $m$ and $N$  parameters ranges  imply a maximum EMGD star mass that lies between $10^6M_{\odot}$ and $10^9M_{\odot}$, where $M_{\odot}$ denotes the solar mass. On the other hand, the lowest EMGD star radius ranges between $10^2R_\odot$ and $ 10^5R_\odot$, where $R_\odot$ denotes the solar radius. 
SU($N$) EMGD dark stars hence present a completely distinct 
experimental 
signature that are quite different of Schwarzschild black hole mergers, due to the subsequent analysis of Table I, as well as Eqs. (\ref{mfka1}).

The peak frequency $f_{\rm max}$, emitted from dark SU($N$) MGD stellar mergers, may be apportioned into a range between 30 $\mu${\rm Hz} and $100$ mHz, at the eLISA experiments \cite{Seoane:2013qna}, whereas the LIGO experiment  can currently detect frequencies between 50 {\rm Hz} and 1 KHz. 
The figures below represent the $N$-$m$ parameter space. First, Fig. 4 represents the $\sigma\to\infty$ GR limit, and Figs. 5-7 take into account the MGD (namely, the EMGD for $k=0$), the EMGD for $k=1$, and the EMGD for $k=2$, with finite brane tension lying in the bound $\sigma \gtrsim  3.2\times10^{-6} \;{\rm GeV^4}$. 
 \begin{figure}[H]
\centering\includegraphics[width=7.2cm]{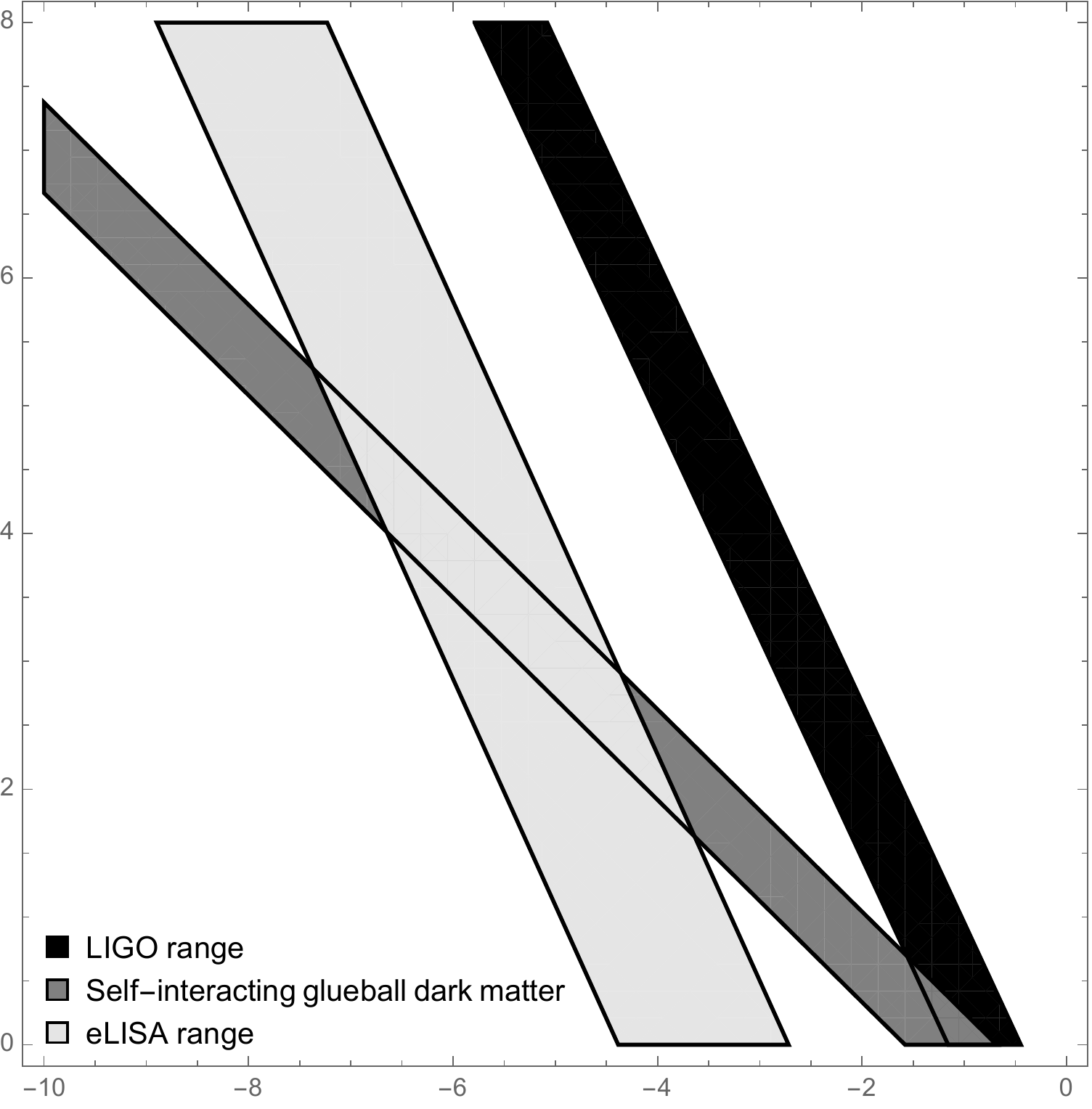}
\caption{The $m$-$N$ parameter space, for the self-interacting glueball dark matter, in the $\sigma\to\infty$ GR limit. The black [light gray] band shows the gravitational wave peak frequency to be detected by the LIGO [eLISA] experiment, whereas the gray band shows the  gravitational wave peak frequency emitted from SU($N$) (Schwarzschild) dark star mergers.}
\end{figure}
 \begin{figure}[H]
\centering\includegraphics[width=7.2cm]{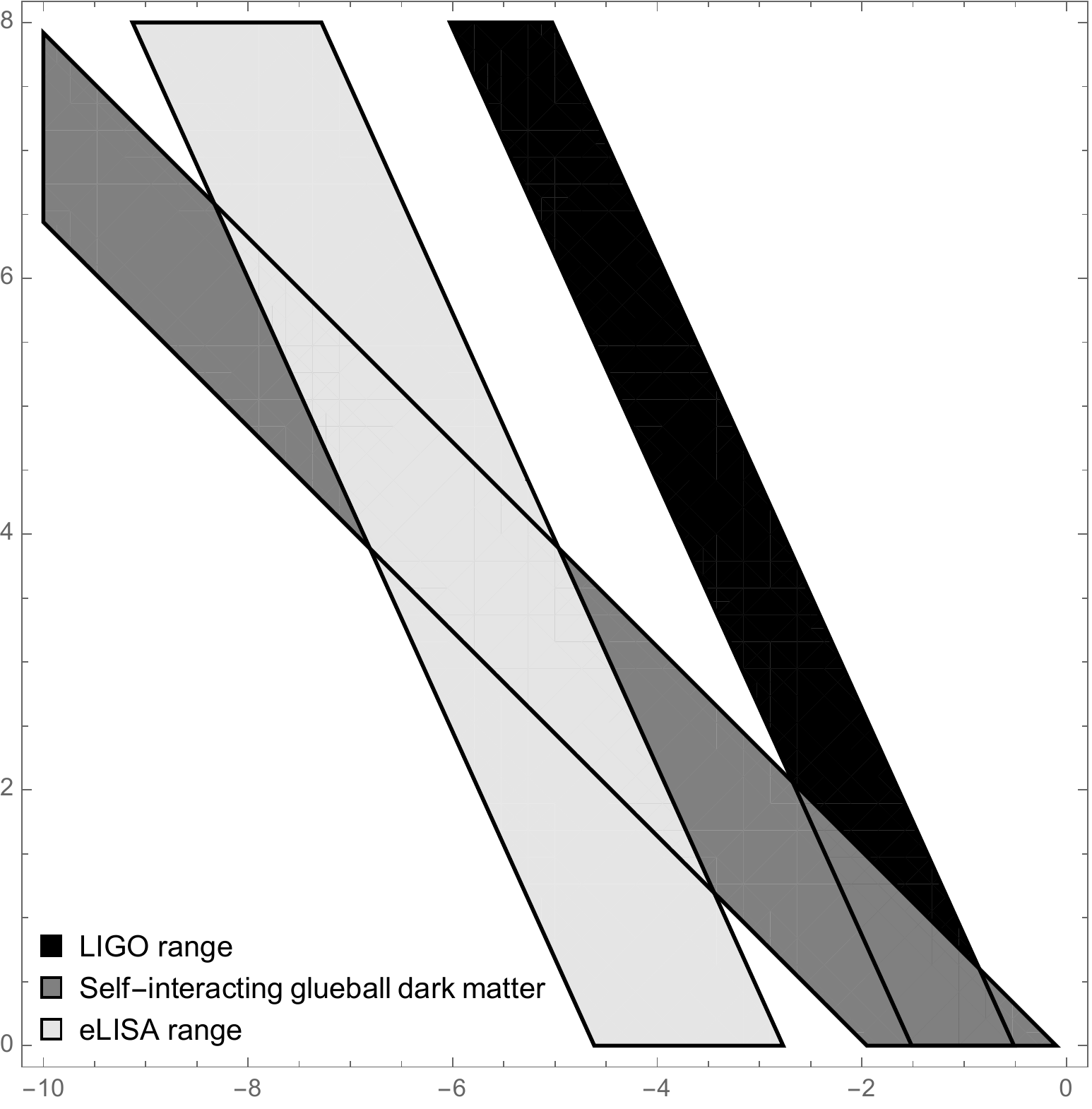}
\caption{The $m$-$N$ parameter space, for the self-interacting glueball dark matter, in the current brane tension bound $\sigma \gtrsim  3.2\times10^{-6} \;{\rm GeV^4}$ \cite{Casadio:2016aum}, for the MGD procedure, corresponding to the EMGD when $k=0$. The black [light gray] band shows the gravitational wave peak frequency to be detected by the LIGO [eLISA] experiment, whereas the gray band shows the  gravitational wave peak frequency emitted from SU($N$) MGD dark star mergers.}
\label{ppppp1}
\end{figure}
 \begin{figure}[H]
\centering\includegraphics[width=7.2cm]{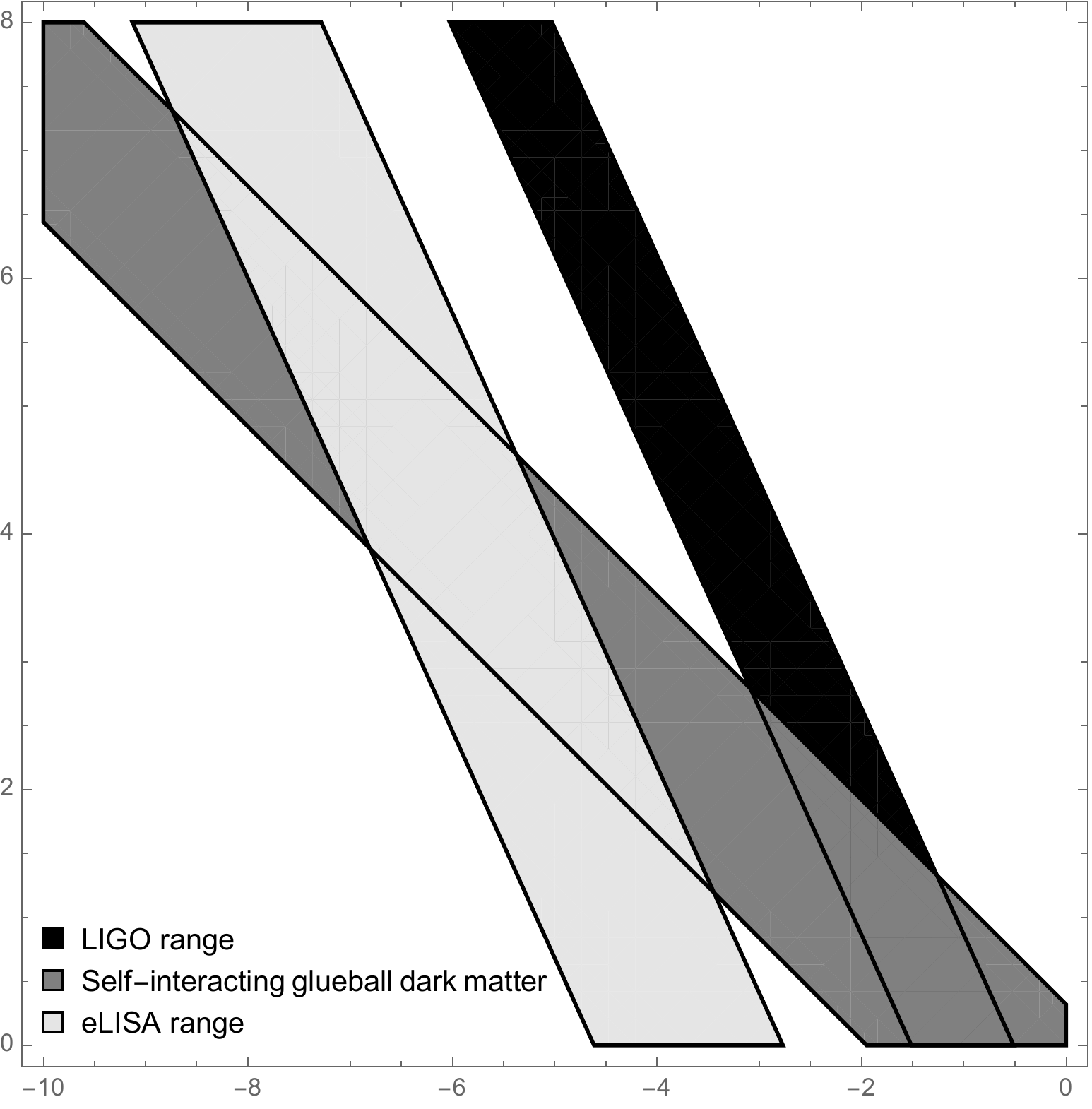}
\caption{The $m$-$N$ parameter space, for the self-interacting glueball dark matter,  in the current brane tension bound $\sigma \gtrsim  3.2\times10^{-6} \;{\rm GeV^4}$ \cite{Casadio:2016aum}, for the EMGD procedure, for $k=1$. The black [light gray] band shows the gravitational wave peak frequency to be detected by the LIGO [eLISA] experiment, whereas the gray band shows the  gravitational wave peak frequency emitted from SU($N$) EMGD dark star mergers.}
\label{ppppp2}
\end{figure}
 \begin{figure}[H]
\centering\includegraphics[width=7.2cm]{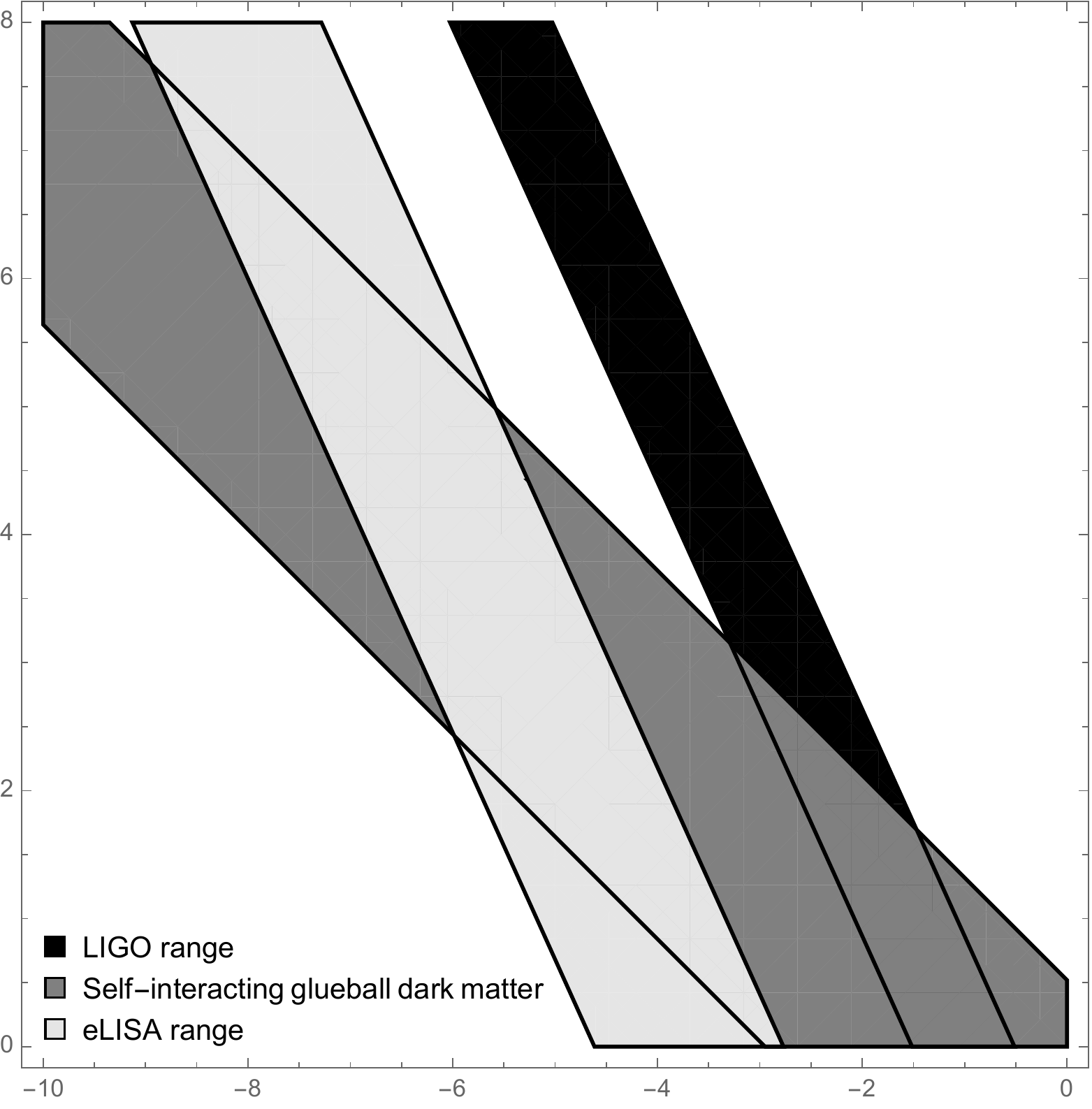}
\caption{The $m$-$N$ parameter space, for the self-interacting glueball dark matter,  in the current brane tension bound $\sigma \gtrsim  3.2\times10^{-6} \;{\rm GeV^4}$ \cite{Casadio:2016aum},  for the EMGD procedure, for $k=2$. The black [light gray] band shows the gravitational wave peak frequency to be detected by the LIGO [eLISA] experiment, whereas the gray band shows the  gravitational wave peak frequency emitted from SU($N$) EMGD dark star mergers.}
\label{ppppp3}
\end{figure}

The bulk Weyl fluid in the membrane paradigm can be, thus, detected,  
in a scenario where hidden sectors to the Standard Model condensate into SU($N$) glueball stars, described by EMGD stellar distributions. Their limiting cases can be further attained, as $k=0$ in  Eq. (\ref{eq:temp_emgd}) corresponds to the standard MGD and, subsequently,  its GR limit $\sigma\to\infty$. In fact, Fig. 7 represents the detection window for gravitational wave radiation emitted by SU($N$) EMGD dark star mergers, for $k=2$ in Eq. (\ref{eq:temp_emgd}), that is wider than in Fig. 6, for $k=1$. Comparing also with Figs. 4 and 5, the bigger the $k$, that drives the extension of the MGD, the wider the window for experimental detection is. In the next section we more precisely analyze our results. 

\section{Conclusions}

SU($N$) hidden gauge sectors were employed to study the scalar glueball dark matter model, when condensation into SU($N$) EMGD dark stars takes part, in the holographic membrane paradigm. Such a process 
then generates stable compact stellar distributions, modelled by the extended MGD. After briefly reviewing the MGD and the EMGD procedures, we focus on the deformation of the metric temporal component by the parameter $k$, in  Eq. (\ref{eq:temp_emgd}). The  
phenomenological upper bound $k\lesssim 4.5$ makes us to study the cases $k=0$ corresponding to the MGD (Fig. 5), whose limit $\sigma\to\infty$, that regards the Schwarzschild solution was also investigated (Fig. 4). For $k=1$ and $k=2$, we illustrate our results in Figs. 6 and 7, respectively. Although the case for $k=1$ resembles a generalized form for the Reissner--Nordstr\"om metric, with tidal charge from a 5D bulk Weyl fluid in Eq. (\ref{eq:k=1}) \cite{CR3}, for $k=2$ the obtained metric (\ref{eq:k=2}) does not take any known form yet. Hence, it may be interesting to study it to estimate the behavior of EMGD solutions in higher orders of $k$, up to the phenomenological bound.
Therefore,  SU($N$) EMGD dark glueball stars present both radius and effective mass that are corrected by a bulk 5D Weyl fluid. Figs. 6 and 7 show that observational signatures evinced from EMGD mergers are more probable to be experimentally detectable at LIGO \cite{Abbott:2007kv} and eLISA project \cite{Seoane:2013qna}. In fact, the window for detection in Figs. 6 and 7 are 
wider than for the  SU($N$) MGD dark glueball stars (Fig. 5), corresponding to the EMGD for $k=0$, and even wider than for SU($N$) Schwarzschild dark glueball stars (Fig. 4). 

It shall be still worth to study the effects of the dynamical corrections and the anomalous dimension effects on the scalar glueballs here studied \cite{Bernardini:2016qit}, as well as glueballs in finite-temperature AdS/QCD models \cite{Miranda:2009uw}, to further refine the spectra of emitted gravitational wave radiation and its consequences in the membrane paradigm. 
Here we proposed that a 5D Weyl fluid, in the membrane paradigm, can induce experimental signatures, at LIGO and eLISA, that are amplified by the EMGD procedure, with respect to the MGD and its $\sigma\to\infty$ GR limit. Hence, our method is alternative to collider data that depends on confirmation at the LHC or future colliders \cite{Deutschmann:2017bth}.

\subsection*{Acknowledgements}
AFS thanks to CAPES, AJFM~is grateful to FAPESP (Grants No.~2017/13046-0 and  2018/00570-5), and RdR~is grateful to CNPq (Grant No. 303293/2015-2),
and to FAPESP (Grant No.~2015/10270-0), for partial financial support. %

\bibliography{bib_DSS1}

\providecommand{\newblock}{}
\begin{thebibliography}{10}
\expandafter\ifx\csname url\endcsname\relax
  \def\url#1{{\tt #1}}\fi
\expandafter\ifx\csname urlprefix\endcsname\relax\def\urlprefix{URL }\fi
\providecommand{\eprint}[2][]{\url{#2}}

\bibitem{Eling:2009sj}
Eling C and Oz Y 2010 {\em JHEP\/} {\bf 02} 069 (\textit{Preprint}
  \eprint{0906.4999})

\bibitem{gub}
Gubser S~S, Klebanov I~R and Polyakov A~M 2002 {\em Nucl. Phys.\/} {\bf B636}
  99--114 (\textit{Preprint} \eprint{hep-th/0204051})

\bibitem{Maldacena:1997re}
Maldacena J~M 1999 {\em Int. J. Theor. Phys.\/} {\bf 38} 1113--1133 [Adv.
  Theor. Math. Phys.2,231(1998)] (\textit{Preprint} \eprint{hep-th/9711200})

\bibitem{Bilic:2015uol}
Bilic N 2016 {\em Phys. Rev.\/} {\bf D93} 066010 (\textit{Preprint}
  \eprint{1511.07323})

\bibitem{hub}
Hubeny V~E 2011 {\em Class. Quant. Grav.\/} {\bf 28} 114007

\bibitem{CR3}
Casadio R, Ovalle J and da~Rocha R 2015 {\em Class. Quant. Grav.\/} {\bf 32}
  215020 (\textit{Preprint} \eprint{1503.02873})

\bibitem{daRocha:2017cxu}
da~Rocha R 2017 {\em Phys. Rev.\/} {\bf D95} 124017 (\textit{Preprint}
  \eprint{1701.00761})

\bibitem{Ovalle:2017wqi}
Ovalle J, Casadio R, da~Rocha R and Sotomayor A 2018 {\em Eur. Phys. J.\/} {\bf
  C78} 122 (\textit{Preprint} \eprint{1708.00407})

\bibitem{Pinzani-Fokeeva:2014cka}
Pinzani-Fokeeva N and Taylor M 2015 {\em Phys. Rev\/} {\bf 91} D4, 044001

\bibitem{maartens}
Maartens R and Koyama K 2010 {\em Living Rev. Rel.\/} {\bf 13} 5
  (\textit{Preprint} \eprint{1004.3962})

\bibitem{Antoniadis:1998ig}
Antoniadis I, Arkani-Hamed N, Dimopoulos S and Dvali G~R 1998 {\em Phys.
  Lett.\/} {\bf B436} 257 (\textit{Preprint} \eprint{hep-ph/9804398})

\bibitem{ovalle2007}
Ovalle J 2017 {\em Phys. Rev.\/} {\bf D95} 104019 (\textit{Preprint}
  \eprint{1704.05899})

\bibitem{covalle2}
Casadio R and Ovalle J 2014 {\em Gen. Rel. Grav.\/} {\bf 46} 1669
  (\textit{Preprint} \eprint{1212.0409})

\bibitem{darkstars}
Ovalle J, Gergely L~Ã and Casadio R 2015 {\em Class. Quant. Grav.\/} {\bf 32}
  045015 (\textit{Preprint} \eprint{1405.0252})

\bibitem{Ovalle:2007bn}
Ovalle J 2008 {\em Mod. Phys. Lett.\/} {\bf A23} 3247--3263 (\textit{Preprint}
  \eprint{gr-qc/0703095})

\bibitem{Casadio:2015jva}
Casadio R, Ovalle J and da~Rocha R 2015 {\em Europhys. Lett.\/} {\bf 110} 40003
  (\textit{Preprint} \eprint{1503.02316})

\bibitem{GCGR}
Shiromizu T, Maeda K~i and Sasaki M 2000 {\em Phys. Rev.\/} {\bf D62} 024012
  (\textit{Preprint} \eprint{gr-qc/9910076})

\bibitem{CoimbraAraujo:2005es}
Coimbra-Araujo C~H, da~Rocha R and Pedron I~T 2005 {\em Int. J. Mod. Phys.\/}
  {\bf D14} 1883--1898 (\textit{Preprint} \eprint{astro-ph/0505132})

\bibitem{Cavalcanti:2016mbe}
Cavalcanti R~T, Goncalves A and da~Rocha R 2016 {\em Class. Quant. Grav.\/}
  {\bf 33} 215007 (\textit{Preprint} \eprint{1605.01271})

\bibitem{Casadio:2016aum}
Casadio R and da~Rocha R 2016 {\em Phys. Lett.\/} {\bf B763} 434--438
  (\textit{Preprint} \eprint{1610.01572})

\bibitem{daRocha:2017lqj}
da~Rocha R 2017 {\em Eur. Phys. J.\/} {\bf C77} 355 (\textit{Preprint}
  \eprint{1703.01528})

\bibitem{CR5}
Casadio R, Ovalle J and da~Rocha R 2014 {\em Class. Quant. Grav.\/} {\bf 31}
  045016 (\textit{Preprint} \eprint{1310.5853})

\bibitem{Juknevich:2009ji}
Juknevich J~E, Melnikov D and Strassler M~J 2009 {\em JHEP\/} {\bf 07} 055
  (\textit{Preprint} \eprint{0903.0883})

\bibitem{Forestell:2016qhc}
Forestell L, Morrissey D~E and Sigurdson K 2017 {\em Phys. Rev.\/} {\bf D95}
  015032 (\textit{Preprint} \eprint{1605.08048})

\bibitem{ArkaniHamed:2008qn}
Arkani-Hamed N, Finkbeiner D~P, Slatyer T~R and Weiner N 2009 {\em Phys.
  Rev.\/} {\bf D79} 015014 (\textit{Preprint} \eprint{0810.0713})

\bibitem{Boddy:2014qxa}
Boddy K~K, Feng J~L, Kaplinghat M, Shadmi Y and Tait T~M~P 2014 {\em Phys.
  Rev.\/} {\bf D90} 095016 (\textit{Preprint} \eprint{1408.6532})

\bibitem{Boddy:2014yra}
Boddy K~K, Feng J~L, Kaplinghat M and Tait T~M~P 2014 {\em Phys. Rev.\/} {\bf
  D89} 115017 (\textit{Preprint} \eprint{1402.3629})

\bibitem{Soni:2016gzf}
Soni A and Zhang Y 2016 {\em Phys. Rev.\/} {\bf D93} 115025 (\textit{Preprint}
  \eprint{1602.00714})

\bibitem{Hartmann:2012wa}
Hartmann B and Riedel J 2012 {\em Phys. Rev.\/} {\bf D86} 104008
  (\textit{Preprint} \eprint{1204.6239})

\bibitem{dadhich}
Dadhich N, Maartens R, Papadopoulos P and Rezania V 2000 {\em Phys. Lett. B\/}
  {\bf 487} 1 (\textit{Preprint} \eprint{hep-th/0003061})

\bibitem{Bernardini:2016qit}
Bernardini A~E, Braga N~R~F and da~Rocha R 2017 {\em Phys. Lett.\/} {\bf B765}
  81--85 (\textit{Preprint} \eprint{1609.01258})

\bibitem{Soni:2016yes}
Soni A and Zhang Y 2017 {\em Phys. Lett.\/} {\bf B771} 379--384
  (\textit{Preprint} \eprint{1610.06931})

\bibitem{Abbott:2007kv}
Abbott B~P {\em et~al.\/} (LIGO Scientific) 2009 {\em Rept. Prog. Phys.\/} {\bf
  72} 076901 (\textit{Preprint} \eprint{0711.3041})

\bibitem{Seoane:2013qna}
Seoane P~A {\em et~al.\/} (eLISA) 2013  (\textit{Preprint} \eprint{1305.5720})

\bibitem{Miranda:2009uw}
Miranda A~S, Ballon~Bayona C~A, Boschi-Filho H and Braga N~R~F 2009 {\em
  JHEP\/} {\bf 11} 119 (\textit{Preprint} \eprint{0909.1790})

\bibitem{Deutschmann:2017bth}
Deutschmann N, Flacke T and Kim J~S 2017 {\em Phys. Lett.\/} {\bf B771}
  515--520 (\textit{Preprint} \eprint{1702.00410})

\end{thebibliography}

\end{document}